\documentclass{ws-mpla}
\usepackage[super]{cite}
\usepackage{graphicx}
\usepackage{color}
\def\nn {\nonumber}

\newcommand{\be}{\begin{equation}}
\newcommand{\ee}{\end{equation}}
\newcommand{\bea}{\begin{eqnarray}}
\newcommand{\eea}{\end{eqnarray}}
\newcommand{\om}{\omega}

\newcommand{\bk}{\bm k}

\begin{document}

\markboth{K. Saha, S. Upadhaya, S. Ghosh}
{A comparative study on two different expressions of Bulk viscosity 
in the PNJL model}

%%%%%%%%%%%%%%%%%%%%% Publisher's Area please ignore %%%%%%%%%%%%%%
\catchline{}{}{}{}{}
%%%%%%%%%%%%%%%%%%%%%%%%%%%%%%%%%%%%%%%%%%%%%%%%%%%%%%%%%%%%%%%%%%%

\title{A comparative study on two different approaches of Bulk viscosity 
in the Polyakov-Nambu-Jona-Lasinio model}

\author{\footnotesize Kinkar Saha$^1$, Sudipa Upadhaya$^2$}

\address{Center for Astroparticle Physics \&
Space Science, Block-EN, Sector-V, Salt Lake, Kolkata-700091, INDIA 
 \\
Department of Physics, Bose Institute,
93/1, A. P. C Road, Kolkata - 700009, INDIA\\
$^1$saha.k.09@gmail.com and $^2$sudipa.09@gmail.com}

\author{Sabyasachi Ghosh}

\address{Department of Physics, University of Calcutta, 92, A. P. C. R
oad, Kolkata - 700009, India\\
sabyaphy@gmail.com}

\maketitle

\pub{Received (Day Month Year)}{Revised (Day Month Year)}

\begin{abstract}
We have gone through a comparative study on two different kind
of bulk viscosity expressions by using a common dynamical model. The 
Polyakov-Nambu-Jona-Lasinio (PNJL) model in the realm of mean-field 
approximation, including up to eight quark interactions for 2+1 flavor
quark matter, is treated for this common dynamics. We have probed 
the numerical equivalence as well as discrepancy of two 
different expressions for bulk viscosity at 
vanishing quark chemical potential. Our estimation of bulk viscosity
to entropy density ratio follows a decreasing trend with temperature,
which is observed in most of the earlier investigations. We have also
extended our estimation for finite values of quark chemical potential. 

\end{abstract}

\ccode{PACS Nos.: 12.38.Aw, 12.38.Mh, 12.39.-x.}

\section{INTRODUCTION}
Ideal hydrodynamics successfully described the experimental data of 
transverse momentum ($p_T$) spectra 
and elliptic flow coefficient $v_2(p_T)$ ~\cite{Ollitrault} for different
hadrons, produced in the heavy ion collisions.
However, one can expect a weakly interacting gas like behavior of the medium,
produced in the heavy ion collisions because of
the asymptotic freedom of Quantum Chromo Dynamics (QCD).
The argument went stronger when one coupled a 2+1 
\cite{Romatschke1,Romatschke10,Song1,Song1a,Dusling1,Bozek,AKC} or 3+1 \cite{Schenke1,Bozek1} 
dimensional ideal hydrodynamical model of Quark Gluon Plasma 
(QGP) phase to a hadron cascade one to properly 
account for the viscous behavior in late hadronic stage \cite{Bass1,Bleicher,Teaney,Ryu,Song,Shen}.
This eventually led to a concept of {\it perfect fluidity}~\cite{Heinz1}, with 
an implication to be a strongly coupled plasma \cite{Gyulassy1,Gyulassy2,Shuryak1}.
\par
In Au+Au and Cu+Cu collisions at Relativistic Heavy Ion Collider (RHIC), 
elliptic flow $v_2$ \cite{Ollitrault,Stocker1,Bozek10,Steinheimer}
is found to be large and not generally consistent with experimental observations upon 
consideration of different types of initial conditions. The absence of proper knowledge
of initial conditions hints about the presence of viscosities in the QGP phase to reproduce
$v_2$ data. It is observed
\cite{Romatschke1,Romatschke10,Song1,Song1a,Heinz3,Song3,Molnar1,Molnar2,Teaney3,Heinz4,Romatschke2}
that both shear and bulk viscosities suppress elliptic flow $v_2$.
However the smallness of this suppression leads to a prediction of very small specific 
shear viscosity value \cite{Kovtun1,Policastro1} for the temperature region probed by 
RHIC or Large Hadron Collider (LHC). 
\par
Bulk viscosity $\zeta$ manifests itself by an addition of a diagonal term $\pi\delta^{ij}$
to stress tensor $T^{ij}$ in the local rest frame. This happens due to local isotropic
deviations from equilibrium. Bulk viscous pressure $\Pi$ becomes proportional to scalar
expansion rate $\theta$ at the location of fluid cell in the Navier-Stoke's approximation,
i.e. $\Pi(x)=-\zeta\theta(x)$. So, expansion is opposed by the bulk term, manifested in
the negative value. Thus, for an isotropically expanding fireball, bulk viscosity reduces the
radial acceleration, thus decreasing radial flow. The picture is somewhat different in
case of shear viscosity. Increase in radial flow by the shear term leads to flatter shape of
particle $p_T$ spectra \cite{Dusling2}, whereas the situation is reverse for bulk viscosity
\cite{Song3,Romatschke3,Song4}. Alongside, in Ref.~\cite{Arnold2} it has been found that for any 
reasonably small running coupling constant bulk viscosity is quantitatively less compared 
to shear viscosity.
\par
A list of references on the microscopic calculations of bulk viscosity coefficient
is not very small~\cite{{Moore1},{Buchel1},{Onuki1},{Paech1},{Gavin1},
{Weinberg1},{Arnold2},{Buchel2},{Prakash1},{Davesne1},{Chen1},{Kharzeev1},
{Kharzeev2},{Fernandez1},{Dobado1},{Marty1},{Xiao1},{Sasaki},{Sasaki2},{G_IFT},{Kadam3},
{Purnendu},{Vinod},{Vinod2},{Santosh},{Sarkar},{Sarkar2},{Kadam},{Kadam2},{Kadam10},
{Sarwar},{Noronha},{Nicola1},{Nicola2},{SG_NISER},{Meyer3}}.
Moore and Saremi \cite{Moore1}
 and Buchel \cite{Buchel1} found that for systems with diverging specific heat near
 transition temperature $T_c$, the relaxation time can blow up with $\frac{\zeta}{s}$ 
still remaining finite. Coming to the expected nature of $\frac{\zeta}{s}$, it should vanish 
for a system of massless non-interacting quanta due to conformal invariance. Presence of 
interactions in the system changes the scenario and leads to deviation from this conformal limit.
 However the deviation remains very small in all regions except near phase transition
 where strong interactions may lead to large correlation length \cite{Moore1,Onuki1,Paech1}.
 Kinetic theory approaches estimate $\zeta$ to be varying as second power in this 
deviation. The technique has been applied in relaxation time approximation \cite{Gavin1}
 as well as for systems of photons radiated by massive particles in equilibrium 
\cite{Weinberg1}, to which leading order QCD results \cite{Arnold2} also agree.
 For strongly coupled $\mathcal{N}$=4SYM theory this variation however
 is taken to be linear in the deviation \cite{Buchel2}. Computation for bulk viscous 
effects has been done by few authors considering hadron gas \cite{Prakash1,Davesne1,Chen1}
as well. Increase of $\frac{\zeta}{s}$ has been observed towards lower temperatures for massive
 pions \cite{Prakash1}, in contrast to a decrease for massless pions \cite{Chen1}. However,
 near phase transition peak like behavior of specific bulk viscosity supports the general
 arguments \cite{Paech1,Kharzeev1,Kharzeev2}, which happens because of long range
 correlations closely related to the chiral symmetry restoration. In presence of 
second order phase transition, $\frac{\zeta}{s}$ is supposed to diverge \cite{Moore1,Buchel1}.
The critical behavior of bulk viscosity in Gross-Neveu and linear sigma models have also 
been explored in Ref.~\cite{Fernandez1} and Ref.~\cite{Dobado1} respectively. 

In the Nambu-Jona-Lasinio (NJL) model, bulk viscosity coefficients $\zeta$ of quark matter
has been calculated by Sasaki et al.~\cite{Sasaki2}, Marty et al.~\cite{Marty1},
Ghosh et al.~\cite{G_IFT} and Deb et al.~\cite{Kadam3}. They have used the expression
of bulk viscosity, which is based on the relaxation time approximation (RTA) 
in kinetic theory approach~\cite{Sasaki2,Marty1,Kadam3} or 
quasi-particle approach of Kubo formalism~\cite{G_IFT}.
Xiao et al.~\cite{Xiao1} has calculated $\zeta$ in the Polyakov-Nambu-Jona-Lasinio (PNJL)
model by using the expression based on the Kubo formalism with sum rule approach.
Using the former and latter approaches, 
Refs.~\cite{Arnold2,Marty1,Sasaki2,G_IFT,Kadam3,Purnendu,Nicola1,Kadam2} and 
Refs.~\cite{Kharzeev1,Kharzeev2,Noronha,Kadam} have respectively estimated $\zeta$
in different other dynamical models. In this context, our interest lies in the study of
both expressions of $\zeta$ under the framework of a common dynamical model and search
some equivalence between two approaches. We have chosen 
Polyakov-Nambu-Jona-Lasinio (PNJL) model as our common dynamics to apply on both
approaches of bulk viscosity calculations. 
\par
The paper is organized as follows. Next section contains formalism part, where a brief
discussion of PNJL model has been mentioned first and then the expression of bulk viscosity
in two approaches are addressed. Next, in result section, the equivalence of two expressions
of bulk viscosity is mainly investigated in numerical point of view for vanishing quark
chemical potential ($\mu$) and then results of finite $\mu$ have also been addressed. 
Finally we have made summary of our work.
%%%%%%%%%%%%%%%%%%%%%%%%%%%%%%%%%%%%%%%%%%%%%%%%%%%%%%%%%%%%%%%%%%%%%%%%%%%%%%%%%%%%%%%%%
%%%%%%%%%%%%%%%%%%%%%%%%%%%%%%%%%%%%%%%%%%%%%%%%%%%%%%%%%%%%%%%%%%%%%%%%%%%%%%%%%%%%%%%%%
\section{Formalism}
Kubo Formalism \cite{Kubo1,Kubo2} relates viscosity coefficients to 
the correlation functions of the energy-momentum tensor. 
For the bulk viscosity coefficient $\zeta$ of QCD matter, the relation is
\be
\zeta=\lim_{\omega\rightarrow 0}\frac{{-\rm Im}G^R(\om,{\vec 0})}{9\omega}
=\lim_{\omega\rightarrow 0}\frac{\pi\rho(\om,{\vec 0})}{9\omega}~,
\label{zeta_Kubo}
\ee
where imaginary part of retarded Green function ${\rm Im}G^R$ and
corresponding spectral density $\rho$, associated with $\zeta$,
are defined as
\be
{-\rm Im}G^R(\om,{\vec 0})=\pi\rho(\om,{\vec 0})
=\int_{0}^{\infty}dt
\int d^3\vec{r} e^{i\omega t}\langle[\theta_\mu^\mu(x),\theta_\mu^\mu(0)]\rangle~.
\ee
The structure of trace of energy-momentum stress tensor $\theta_\mu^\mu$ for QCD in low-energy theorems 
at finite temperature and density has been addressed in Ref.~\cite{Kapusta2}.
Using standard Kramers-Kronig relation alongwith some basic thermodynamics and
considering the quark and gluon components,
we get~\cite{Kharzeev2,Kharzeev1,Xiao1}
\bea
-G^R(0,{\vec 0})&=&2\int_0^\infty\frac{\rho(u,\vec{0})}{u}du 
=-6\Big( -f_\pi^2 M_\pi^2-f_K^2 M_K^2\Big) +16|\epsilon_v| +Ts\left( \frac{1}{c_s^2}-3\right) 
\nn\\
&& +\left( \mu\frac{\partial}{\partial\mu}-4\right) T^5\frac{\partial(\frac{P}{T^4})}{\partial T}
+\left( T\frac{\partial}{\partial T}+ \mu\frac{\partial}{\partial\mu}-2\right)
 \big \langle m\bar{q}q\big \rangle_T 
~.
\nn\\
\label{sumrule}
\eea

To extract bulk viscosity $\zeta$, an ansatz is to be made for the spectral density $\rho$.
 Since the divergent contribution is deducted in the definition of quantities in right hand
 side of Eq.(\ref{sumrule}) (realized in terms of the first term in right hand side),
 so the high frequency perturbative continuum \cite{Fujii} 
$\rho(u)\sim\alpha_s^2u^4$ is not included in left hand side of the same. In low 
frequency region, the following ansatz is chosen,
\begin{equation}
 \frac{\rho(\omega,\vec{0})}{\omega}=\frac{9\zeta}{\pi}\frac{\omega_0^2}{\omega_0^2+\omega^2}
\label{rho-ansatz}
\end{equation}
such that it satisfies Eq.(\ref{zeta_Kubo}), where $\omega_0$ is the mass scale corresponding to 
region of validity of perturbation theory. At $\omega_0>>T$, it is expected that the spectral 
density will become perturbative and temperature-independent. So we obtain,
\bea
 \zeta &=& \frac{1}{9\omega_0}\Bigg[ Ts \left( \frac{1}{c_s^2}-3 \right) 
+ \left( \mu\frac{\partial}{\partial\mu}-4 \right) T^5\frac{\partial(\frac{P}{T^4})}{\partial T}
\nn\\
&&+\left( T\frac{\partial}{\partial T}+
\mu\frac{\partial}{\partial\mu}-2\right)  \big \langle m\bar{q} q\big \rangle_T 
+6\left( f_\pi^2 M_\pi^2 +f_K^2 M_K^2 \right) +16|\epsilon_v|
\Bigg]
\label{zetafinal}
\eea
which is the final expression for bulk viscosity coefficient, $\zeta$, based on QCD sum rule.

Next, we will come to another possible expression of $\zeta$, which is based on 
the relaxation time approximation (RTA) in kinetic theory approach~\cite{Gavin1,Purnendu}. 
Exactly same expression can also be derived from the quasi-particle approach of Kubo formalism,
where spectral density of $\zeta$ can be expressed in terms of one-loop diagram~\cite{Nicola1,Nicola2,G_IFT}.
The idea of quasi-particle picture is introduced via inclusion of finite thermal width ($\Gamma$)
in the internal lines of one-loop diagram.
Instead of going to their detail derivation, which one can see in Refs.~\cite{Gavin1,Purnendu} 
for RTA method and in Refs.~\cite{Nicola1,G_IFT} for quasi-particle approach of
Kubo formalism, let us come to their final expression for $\zeta$.
For vanishing quark chemical potential ($\mu=0$), the RTA expression is
\be
\zeta = \frac{12}{T} \int \frac{d^3\bk}{(2\pi)^3} 
\, \frac{ f_{\Phi} \left[1 - f_{\Phi}\right]}{(E_{k})^2 \, \Gamma} 
 \left[ \left(\frac{1}{3} - c_s^2\right) \bk^2 
- c_s^2 \frac{d}{d\beta^2} \left( \beta^2m^2 \right) \right]^2 ~,
\label{zeta_QP}
\ee
where the effect of Polyakov loop is absorbed
into the PNJL distribution function designated by $f_{\Phi}$ \cite{Saumen1,Sudipa1}.
To get the total bulk viscosity, we will have to take the
sum over 2+1 flavors.
Here, $E_{k}=\{\bk^2+m^2\}^{1/2}$ is energy of quark and 
$\Gamma=1/\tau$ is its thermal width, which inversely determine 
its relaxation time $\tau$ in the medium.
\par
Here in this work, we compute the 
bulk viscous coefficient using the methodologies of LET and RTA. 
For our studies,
we use the framework of the 2+1 flavor PNJL model
taking upto eight quark interaction terms. 
This model was developed by addition of Polyakov loop to the 
NJL model \cite{YNambu1,YNambu2,Kunihiro1,Kunihiro2,Vogl1,Klevansky1,Buballa1}.
An insight into its formalism
and recent developments can be found in 
Refs.~\cite{Ghosh1,Swagato,Ratti1,Ratti2,WFu1,WFu2,Bhattacharyya1,Bhattacharyya3,
Bhattacharyya4,Bhattacharyya5,Bhattacharyya6,Bhattacharyya7,Bhattacharyya8,Bhattacharyya2,Ghosh2,VDM}.
%%%%%%%%%%%%%%%%%%%%%%%%%%%%%%%%%%%%%%%%%%%%%%%%%%%%%%%%%%%%%%%%%%%%%%%%

%%%%%%%%%%%%%%%%%%%%%%%%%%%%%%%%%%%%%%%%%%%%%%%%%%%%%%%%%%%%%%%%%%%%%%%%%%%%%%%%%%%%%%%%%%%%%%%%
\section{Results and Discussions}
%%%%%%%%%%%%%%%%%%%%%%%%%%%%%%%%%%%%%%%%%%%%
\begin{figure}   %[!htb]
\begin{center}
 {\includegraphics[height=6cm,width=9.8cm,angle=0]
{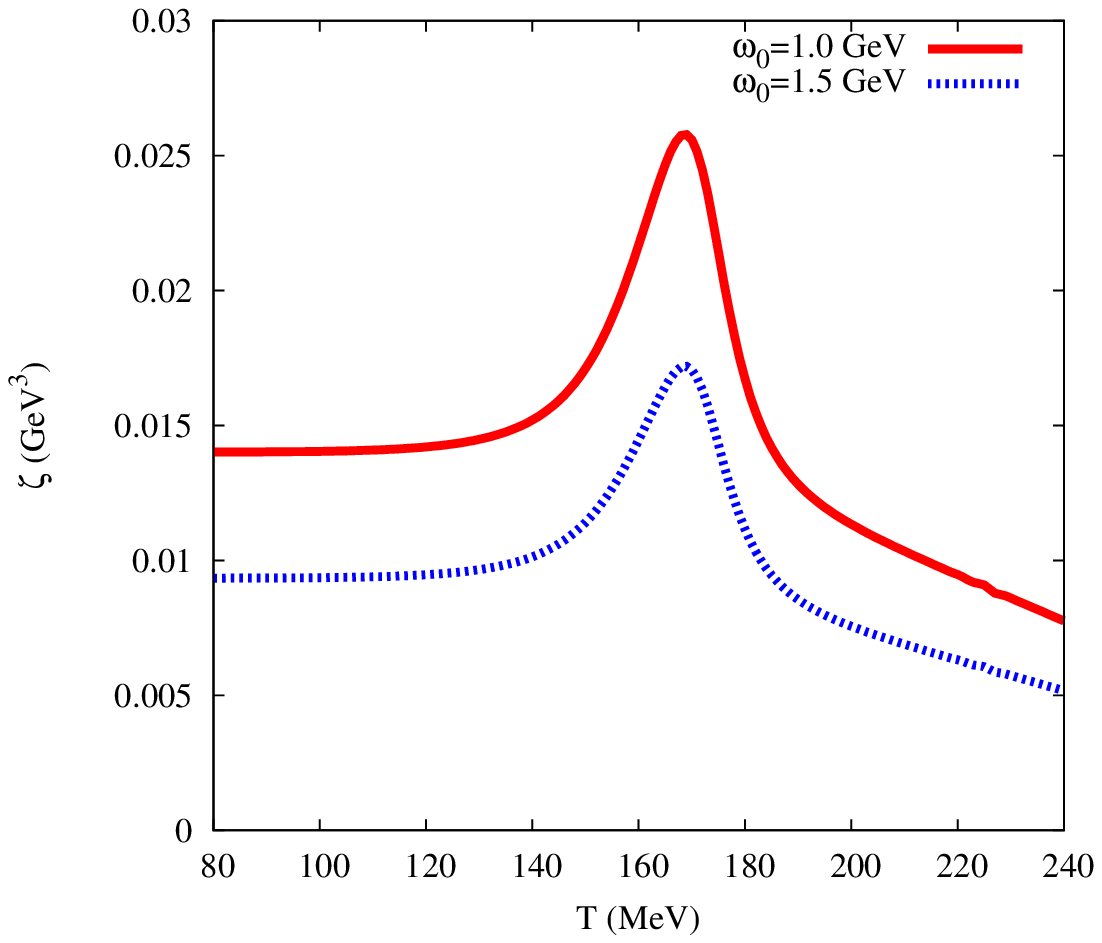}}
{\includegraphics[scale=0.35]{z_QP.eps}}
 \caption{(color online) $\zeta$ as a function of temperature at vanishing
chemical potential. Results of upper and lower panels are obtained
from Eqs.~(\ref{zetafinal}) and (\ref{zeta_QP}), which are based on sum rule
and quasi-particle approaches of Kubo relation (RTA in Kinetic theory) respectively.} 
\label{zetaT}
\end{center}
\end{figure}
%%%%%%%%%%%%%%%%%%%%%%%%%%%%%%%%%%%%%%%%%%%%%%%%%%%%%%%%%%%%%%%%%%%%%%%%
%%%%%%%%%%%%%%%%%%%%%%%%%%%%%%%%%%%%%%%%%%%%%%%%%%%%%%%%%%%%%%%%%%%%%%%%%
\begin{figure}
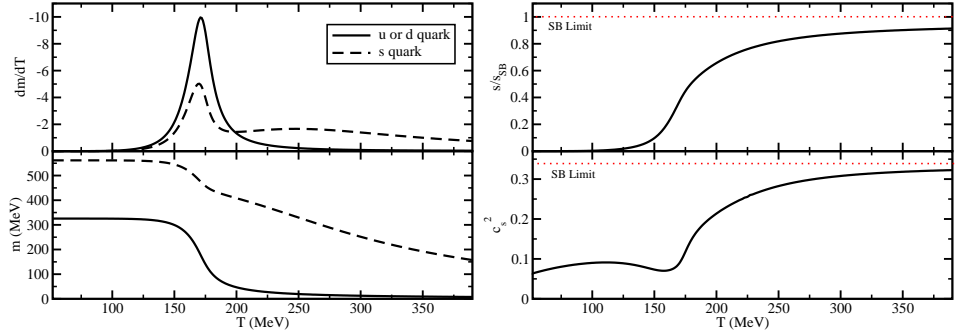
  %[!htb]
\begin{center}
{\includegraphics[scale=0.26]{dM_T.eps}}
{\includegraphics[scale=0.26]{cs2_T.eps}}
\caption{(color online) Left: $T$ dependence of masses $m$ (lower panel) 
and $\frac{dm}{dT}$ (upper panel) of $u$ or $d$ (solid line) and $s$ (dash line) quarks.
Right: $T$ dependence of normalized entropy density $s/s_{_{SB}}$ (upper panel)
and square of speed of sound $c_s^2$ (lower panel), where SB limits of $s$ 
($s=s_{_{SB}}$) and $c_s^2$ ($c_s^2=1/3$) are marked by red dotted lines.}
\label{cs2} 
\end{center}
\end{figure}
%%%%%%%%%%%%%%%%%%%%%%%%%%%%%%%%%%%%%%%%%%%%%%%%%%%%%%%%%%%%%%%%%%%%%%%%%
%%%%%%%%%%%%%%%%%%%%%%%%%%%%%%%%%%%%%%%%%%%%%%%%%%%%%%%%%%%%%%%%%%%%%%%%
\begin{figure}  %[!htb]
\begin{center}
 {\includegraphics[height=9.4cm,width=6.4cm,angle=270]
{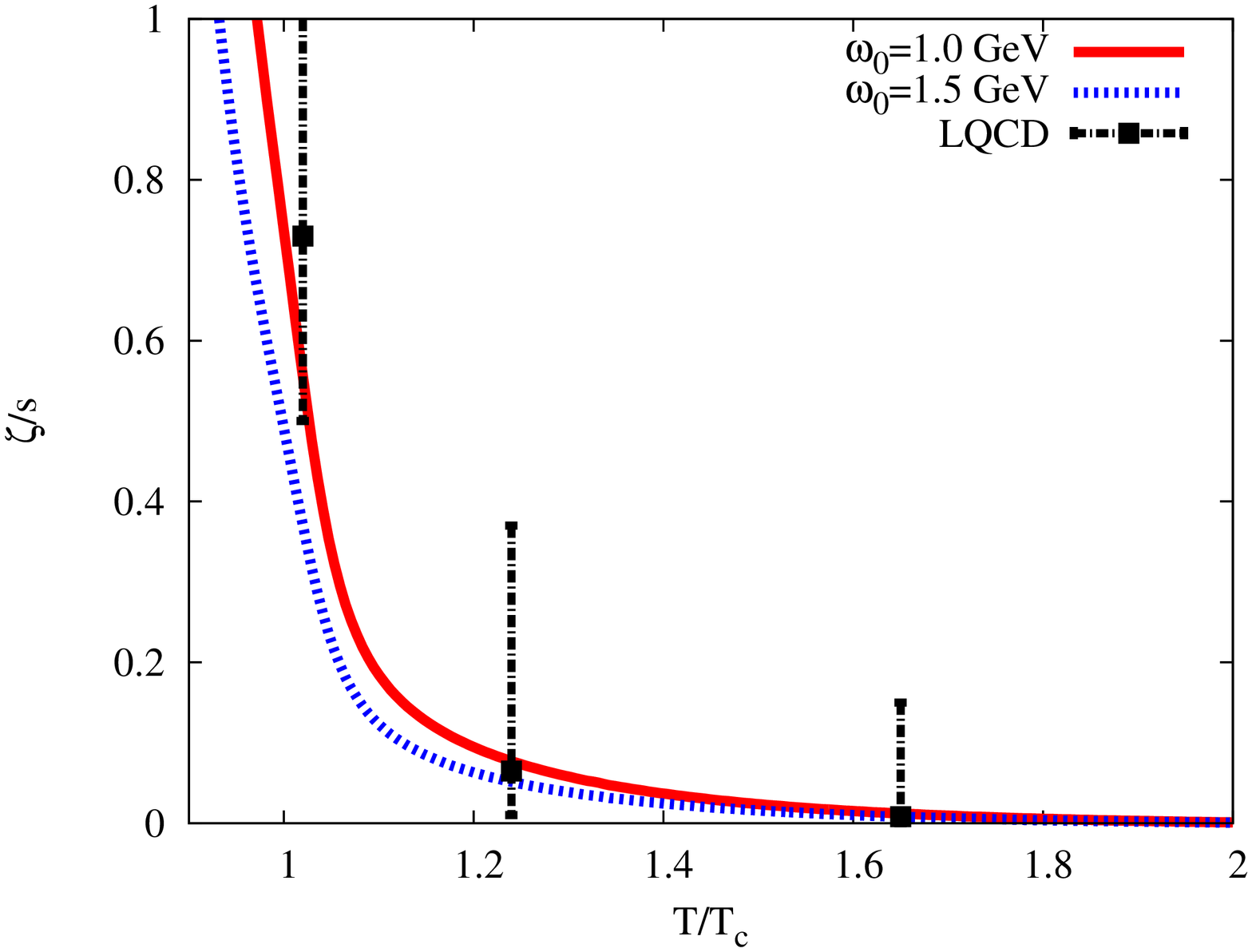}}
{\includegraphics[scale=0.35]{zs_QP.eps}}
\caption{(color online) $\frac{\zeta}{s}$ as a function of temperature at vanishing
chemical potential in the sum rule (upper panel) and quasi-particle (lower panel)
approaches. For comparison, Lattice-QCD values of $\frac{\zeta}{s}$ (squares with error bars) 
from Ref.~\cite{Meyer3} are pasted.} 
\label{zetabysT}
\end{center}
\end{figure}
%%%%%%%%%%%%%%%%%%%%%%%%%%%%%%%%%%%%%%%%%%%%%%%%%%%%%%%%%%%%%%%%%%%%%%%%
%%%%%%%%%%%%%%%%%%%%%%%%%%%%%%%%%%%%%%%%%%%%%%%%%%%%%%%%%%%%%%%%%%%%%%%%
\begin{figure}  %[!htb]
\begin{center}
 {\includegraphics[height=8cm,width=6.0cm,angle=270]
{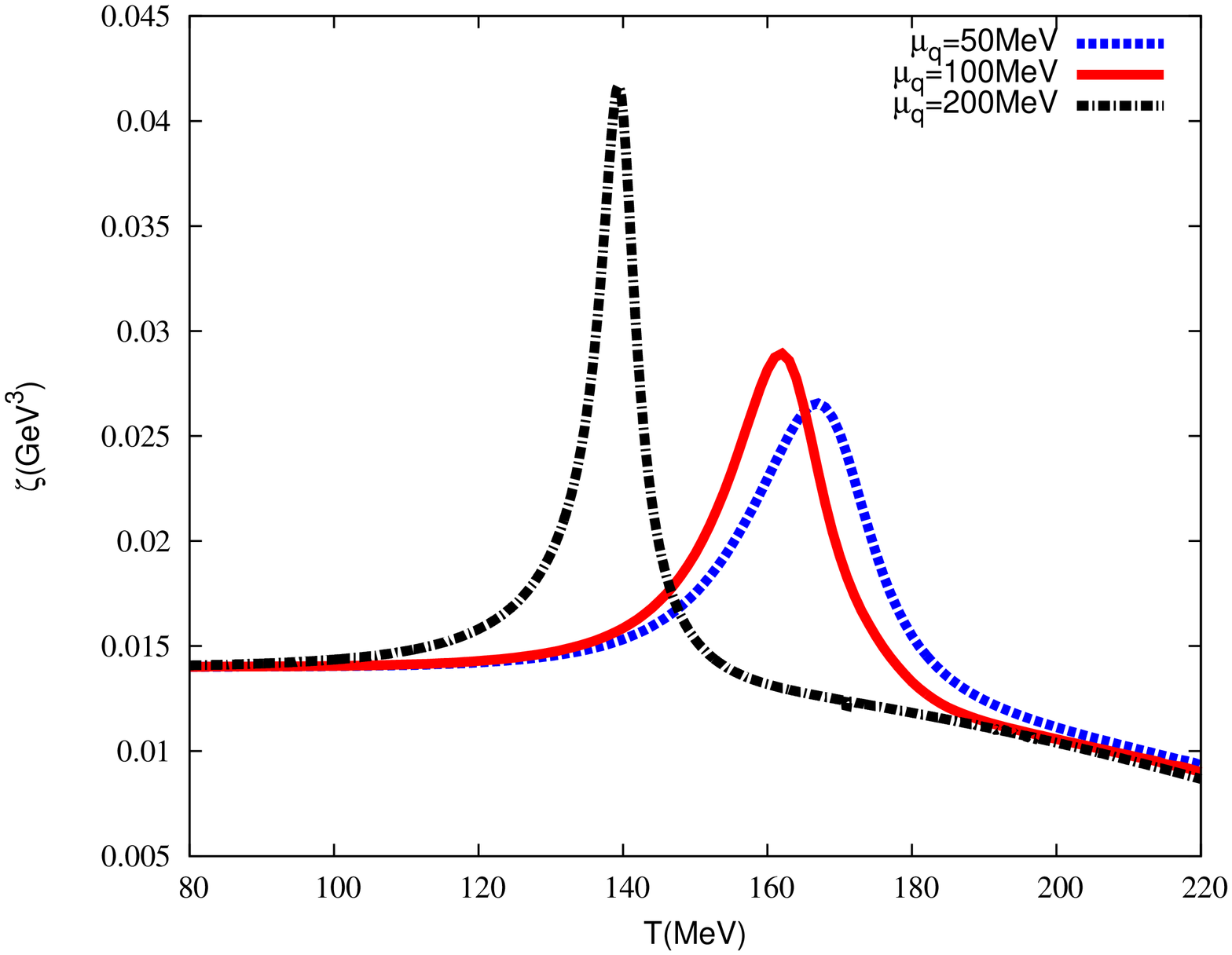}}
 {\includegraphics[height=8cm,width=6.0cm,angle=270]
 {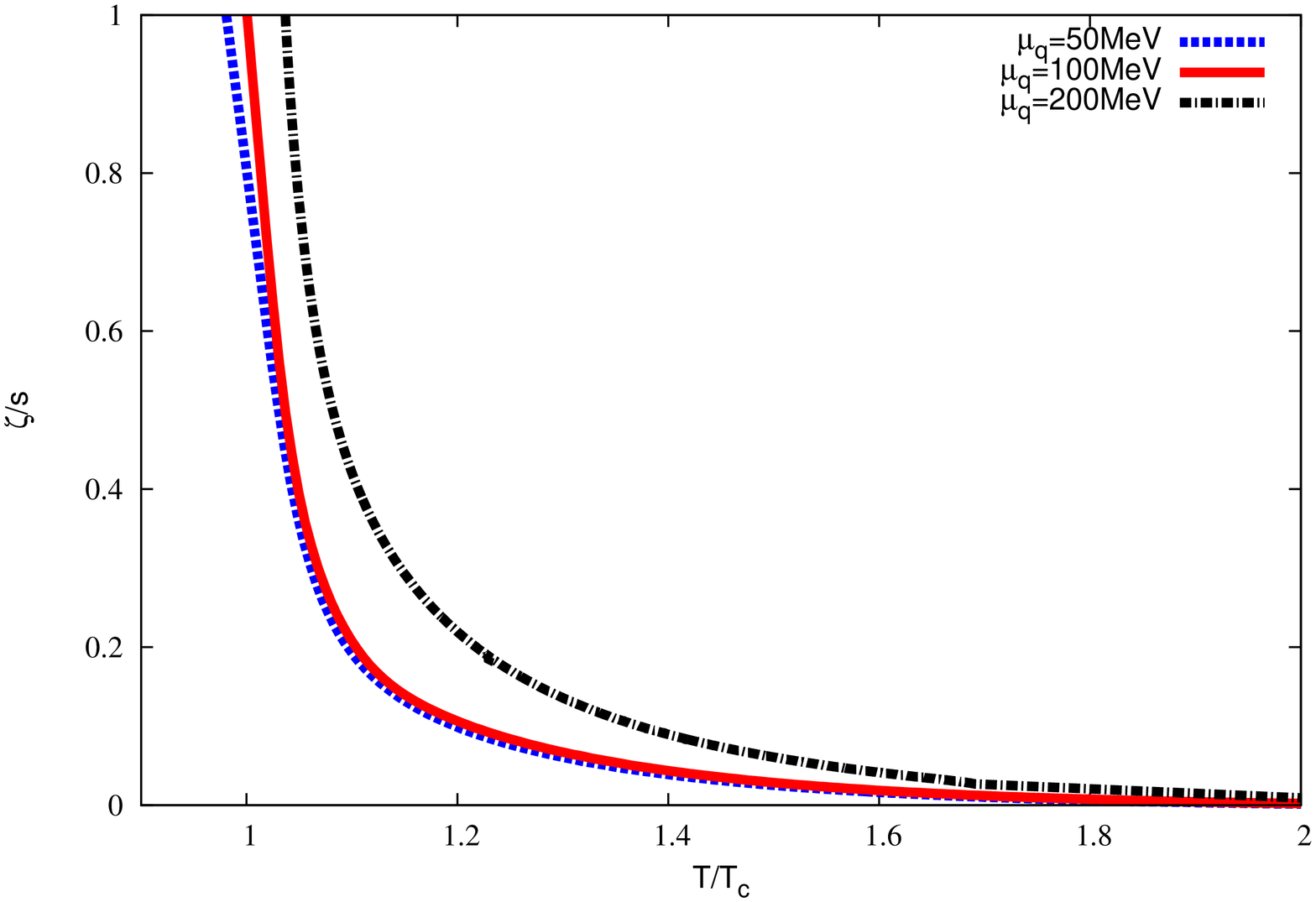}}
\caption{(color online) $\zeta$ (upper panel) and $\frac{\zeta}{s}$ (lower panel) 
as a function of temperature for non-zero quark
chemical potential.} 
\label{zetaTmuq50to200}
\end{center}
\end{figure}
%
%%%%%%%%%%%%%%%%%%%%%%%%%%%%%%%%%%%%%%%%%%%%%%%%%%%%%%%%%%%%%%%%%%%%%%%%
Let us start our discussion from numerical values of $\zeta$, 
obtained from the Eq.~(\ref{zetafinal}) at $\mu=0$. These are 
plotted in the upper panel of Fig.~(\ref{zetaT}) for two different values of $\om_0$,
which inversely control the numerical strength of $\zeta$. Each of them
exhibit visible peak structures near the transition temperature 
$T_c \sim$169 MeV. The temperature dependence of quantities like
$\epsilon - 3P$ and $\big \langle m\bar{q} q\big \rangle_T$ are collectively
responsible for exploring this peak structure. The former quantity is
closely related with speed of sound $c_s$. 
%as disclosed by Eq.~(\ref{cs2_e3P}).
The deviation of $c_s^2$ from its limiting value, $c_s^2=1/3$, measures
the violation of conformality, which is basically determined from 
the term $\Big(\frac{1}{3} -c_s^2\Big)$, associated with the former quantity.
The second quantity $\big \langle m\bar{q} q\big \rangle_T$ vanishes in
the massless limit and exposes a conformally symmetric medium. Hence,
non-zero values of both quantities make a link between the non-zero values
of $\zeta$ and violation of conformality. The peak structure of $\zeta$ near
$T_c$ indicates maximum breaking of conformal symmetry.

Next, let us come to the results of $\zeta(T, \mu=0)$ from the Eq.~(\ref{zeta_QP}),
which are plotted in the lower panel of Fig.~(\ref{zetaT}). Similar to $\om_0$ of
Eq.~(\ref{zetafinal}), $\Gamma=1/\tau$ in Eq.~(\ref{zeta_QP})
inversely controls the numerical strength of $\zeta$. We have chosen two
certain values of $\Gamma$ or $\tau$, which can approximately reproduce 
the $\zeta$ in same order of magnitude as done by $\om_0=1$ and $1.5$ GeV 
in earlier case. On the basis of this approximate matching, we may get an equivalence 
between two parameters:
\bea
\tau=4~ {\rm fm~or,}~\Gamma=0.049~{\rm GeV}~~~&\equiv ~~~&\om_0=1.5~{\rm GeV}
\nn\\
\tau=7~ {\rm fm~or,}~\Gamma=0.028~{\rm GeV}~~~&\equiv ~~~&\om_0=1~{\rm GeV}~,
\eea
which are coming from two different approaches.
In this context, the reader should focus only on the peak structures near the transition temperature
in both approaches because their numerical strength (peak strength) are basically
matched. Although, the numerical values of $\zeta$ at temperatures away from $T_c$
are quite different in two approaches. The possible reason is discussed in the 
next paragraph.

One of the major difference between the two expressions of $\zeta$ is that 
the conformal breaking terms are folded by (PNJL) distribution function of quark
in the RTA expression (\ref{zeta_QP}) but it is absent in the LET expression (\ref{zetafinal}).
Now, this (PNJL) distribution function of quark always has a suppressing 
effect at low $T$ after the folding or integration operation.
So, this is the mathematical reason for lower values of $\zeta(T)$ at $T<T_c$ in
RTA approach than that of sum rule approach. One may expect this kind of
folding by the quark distribution function in the LET expression of $\zeta$ 
if one consider some different kind of ansatz for this spectral density, 
which may contain this kind of folding. 
In ref.~\cite{Kharzeev2}, it is distinctly mentioned that 
a possible uncertainty in this sum rule approach may be appeared 
from this ansatz for the spectral density. 

In the RTA approach, the terms of conformal symmetry breaking are 
$\left(\frac{1}{3} - c_s^2\right)$ and $\frac{d}{d\beta^2} \left( \beta^2m^2_Q \right)$.
Similar to earlier case, one can check again that both of the terms
vanish in the massless limit. The role of 
$\frac{d}{dT}\big \langle m\bar{q} q\big \rangle_T$ in the sum rule approach
is equivalently played by the term, 
$\frac{d}{d\beta^2} \left( \beta^2m^2 \right)=m^2 + Tm\frac{dm}{dT}$,
of the quasi-particle approach. To understand the temperature dependence of
these quantities, we have plotted quark masses (lower panel)
and their temperature derivative (upper panel) in the left-hand side 
of Fig.~(\ref{cs2}). 
The peak structures
of $\zeta$ are built by the peak structures of $\frac{dm}{dT}$, 
which basically represent the transition between broken
and restored phases of chiral symmetry. In the right hand side of Fig.~(\ref{cs2}),
the entropy density $s$ (upper panel) and the square of speed of 
sound $c_s^2$ (lower panel) are plotted against temperature axis.
Entropy density at high $T$ domain
reach to Stefan-Boltzmann limit(SB-limit), $s_{_{SB}}=\frac{19\pi ^2}{9}T^3$, 
under 2+1 flavor consideration. The $s/s_{_{SB}}$ is plotted in the 
upper-right panel of Fig.~(\ref{cs2}), where red dotted line is denoting the 
SB limit $s/s_{_{SB}}=1$. Being inversely related with $\frac{ds}{dT}$,
$c_s^2$ exhibits a dip near $T_c$, where rate of increment of $s(T)$ is larger.
Hence one naturally gets a peak structure in 
$\left(\frac{1}{3} - c_s^2\right)$ near $T_c$, which contributes in the peak
structure of $\zeta$. SB limit of $c_s^2$ is marked by red dotted line in the figure.
To display the approach of the thermodynamical quantities towards
their SB limit, we have plotted Fig.~\ref{cs2} up to $T=400$ MeV 
to cover the high $T$ region.

Next, upper and lower panels of Fig.~(\ref{zetabysT}) show the temperature dependence
of specific bulk viscosity $\zeta/s$, obtained from two different approaches with the 
same parameters, taken for $\zeta(T)$ in Fig.~(\ref{zetaT}). In both approaches,
we observe decreasing nature of $\zeta/s(T)$ in the region of $T>T_c$.
Since $\zeta$ from sum rule approach is quite larger than that from quasi-particle 
approach at low temperature ($T<T_c$), so $\zeta/s$ increases to large numbers
as one decreases the temperature at low temperature domain ($T<T_c$). 
That is why the low $T$ results have not been shown in the upper panel of Fig.~(\ref{zetabysT}).
In the lower panel of Fig.~(\ref{zetabysT}), results of quasi-particle approach show
peak structures at $T_c$ and hence, a discrepancy between the quantitative nature
of $\zeta/s (T<T_c)$ in two approaches are observed. 
This is because the $\zeta(T<T_c)$ from quasi-particle approach 
is quite smaller than that from sum rule approach.
The $\zeta/s$ at $T\geq T_c$ are 
more or less same in both approaches and quite satisfactorily 
agree with LQCD results, obtained
in Ref.~\cite{Meyer3}.
At high temperature domain, $\frac{\zeta}{s}$ in both approaches are tending to be zero.
It indicates that both methodologies are obeying the general fact
of high temperature QCD, which approaches towards the massless limit
to attain its conformal symmetric behavior.
%%%%%%%%%%%%%%%%%%%%%%%%%%%%%%%%%%%%%%%%%%%%%%%%%%%%%%%%%%%%%%%%%%%%%%%%
%%%%%%%%%%%%%%%%%%%%%%%%%%%%%%%%%%%%%%%%%%%%%%%%%%%%%%%%%%%%%%%%%%%%%%%%

We have extended our investigation for finite quark chemical potential by
using Eq.~(\ref{zetafinal}) of sum rule approach in Kubo framework.
In the upper panel of Figure(\ref{zetaTmuq50to200}), $\zeta$ is plotted for three different values of 
quark chemical potential $\mu_q$= 50, 100, 150 MeV keeping charge and strangeness 
chemical potentials ($\mu_Q$ and $\mu_S$ respectively) fixed at zero and $\omega_0~=~1$GeV. 
It is seen that with increase of $\mu_q$, peaks of $\zeta$ are shifted towards
lower $T$ with higher numerical strength.
For different values of chemical potential, the transitions take place at 
different temperatures in the QCD phase diagram, which causes
shifts in the peak positions of bulk viscosity. 
\par
The lower panel of Fig.~(\ref{zetaTmuq50to200}) shows the results for 
$\frac{\zeta}{s}$ as a function of $T$ considering same set of values of 
chemical potentials, where we again see the increasing nature of $\frac{\zeta}{s}$
with increasing of $\mu$. Similar trend was observed for shear viscosity to
entropy density ratio in earlier studies~\cite{Sudipa1}, based on the same PNJL
dynamics.
\begin{table}  %[h]
%\begin{center}
\tbl{List of estimated values of $\zeta/s$ in earlier works at hadronic ($2^{\rm nd}$ column)
and quark ($3^{\rm rd}$ column) temperature domain. Dynamics with references and 
nature of temperature dependence are described in $1^{\rm st}$ and $4^{\rm th}$ columns respectively.}
%\begin{tabular}{|c|c|c|c|}
{\begin{tabular}{@{}cccc@{}} \toprule
%\hline
Dynamics$^{\rm References}$ &  $T ~\leq~T_c$  & $T ~\geq~T_c$ & Nature of $T$ dependence  \\
%\hline
%\hline
%\underline{Kubo formalism with} & & & \\
%\underline{sum rule approach:} & & & \\
\colrule
LQCD~\cite{Meyer3,Kharzeev1,Kharzeev2} & - & $\approx 1-0$ & Decreasing \\
%& & \\
HRG~\cite{Noronha} & $\approx 0.02-0.003$ & - & Decreasing \\
HRG + HS~\cite{Noronha} & $\approx 0.02-0.1$ & - & Increasing \\
%& & \\
HRG~\cite{Kadam} & $\approx 0.01-0.035$ & - & Increasing \\
HRG + HS~\cite{Kadam} & $\approx 0.025-0.12$ & - & Increasing \\
%& & \\
\hline
\hline
%\underline{RTA in Kinetic theory} & & & \\
%\underline{approach/ Quasi-particle} & & & \\
%\underline{ approach of Kubo formalism:} & & & \\
HTL~\cite{Arnold2} & - & $\approx 0.002-0.001$ & Decreasing \\
%& & \\
NJL~\cite{Marty1} & $\approx 0.9-0.02$ & $\approx 0.02-0.002$ & Decreasing \\
%& & \\
NJL~\cite{Sasaki2} & $\approx 1.7-0.13$ & $\approx 0.13-0.005$ & Decreasing \\
%& & \\
NJL~\cite{G_IFT} & - & $\approx 0.1-0.01$ & Decreasing \\
%& & \\
NJL~\cite{Kadam3} & $\approx 0.61-0.11$ & $\approx 0.11-0.004$ & Decreasing \\
%& & \\
LSM~\cite{Purnendu} & $\approx 0.61-0.11$ & $\approx 0.11-0.004$ & Decreasing \\
%& & \\
Unitarization~\cite{Nicola1} & $\approx 0.04-0.027$ & - & Decreasing \\
%& & \\
HRG~\cite{Kadam2} & $\approx 0.15-0.025$ & - & Decreasing \\
%& & \\
%\hline
\botrule
\end{tabular}
\label{tab} }
%\end{center}
\end{table}

To understand our results with respect to earlier works, we have made the table~(\ref{tab}),
which contain approximate numerical values of $\zeta/s$, extracted from some earlier 
works~\cite{Kharzeev1,Kharzeev2,Noronha,Kadam,Arnold2,Marty1,Sasaki2,G_IFT,Kadam3,Purnendu,Nicola1,Kadam2}.
The $1^{\rm st}$ column contains the information of
dynamics for different works and next, the values of $\zeta/s$
at hadronic and quark temperature domain for vanishing quark or baryon 
chemical potential are put in $2^{\rm nd}$ and $3^{\rm rd}$ columns respectively. 
The nature of the function,
$\zeta/s(T)$ for different estimations are mentioned in the last column. We have also
divided the table into two part, where references of upper part have used the expression
of $\zeta$, based on sum rule approach of Kubo formalism and the references of 
lower part have used the RTA or quasi-particle expression of $\zeta$. Now most of the works
observed decreasing nature of $\zeta/s(T)$ in both
$T<T_c$~\cite{Noronha,Nicola1,Kadam2,SG_NISER} and
$T>T_c$~\cite{Meyer3,Kharzeev1,Kharzeev2,Arnold2,G_IFT} regime
as well as in entire 
range of $T$~\cite{Marty1,Sasaki2,Kadam3,Purnendu,Dobado1}. Our results, based on the Kubo
expression of sum rule approach, are supporting this decreasing nature of $\zeta/s(T)$.
In numerical point of view, one can relate our results to the divergence nature of 
$\zeta/s$ near $T_c$, as observed in Refs.~\cite{Meyer3,Kharzeev1,Kharzeev2}. 
Now, on the other hand, the RTA expression is exhibiting a peak structure 
in the plot of $\zeta/s$ vs $T$. This kind of peak structure is also observed
in the Linear Sigma Model (LSM) calculations~\cite{Purnendu,Dobado1} as well as
in NJL model calculations~\cite{G_IFT}. We may assume indication of similar kind
of peak structure from the increasing nature of $\zeta/s(T<T_c)$, observed 
in Refs.~\cite{Kadam,Noronha}, whose calculations are based on the
Hadron Resonance Gas (HRG) model without~\cite{Kadam} and 
with~\cite{Kadam,Noronha} Hagedorn States (HS). So our investigations on
two different expression of $\zeta$ in PNJL model is revealing a qualitative
discrepancy between the $\zeta/s(T<T_c)$ of two approaches, because of quantitative
differences in their $\zeta(T<T_c)$. However, our $\zeta/s(T > T_c)$ are qualitatively
very similar and
for present set of values ($\om_0=1,~1.5$ GeV and $\Gamma=0.049,~0.099$ GeV) in both
approaches match quite satisfactorily with results, 
obtained by Refs.~\cite{Meyer3,Kharzeev1,Kharzeev2}. 
%%%%%%%%%%%%%%%%%%%%%%%%%%%%%%%%%%%%%%%%%%%%%%%%%%%%%%%%%%%%%%%%%%%%%%%%%%%%%%%
\section{Summary}
In the framework of PNJL model with 2+1 flavor consideration, we have gone
through a numerical investigation of two different expressions for bulk viscosity,
based on two different approaches. One is Kubo formalism in sum rule approach
and another is the quasi-particle approach of Kubo framework or
the relaxation time approximation in kinetic theory approach. We have 
computed the temperature dependence of $\zeta$ and $\zeta/s$ for vanishing quark 
chemical potential in both approaches, where peak structure in $\zeta$ near $T_c$ 
and the decreasing nature of $\zeta/s(T)$ (particularly at $T\geq T_c$) are their common
numerical outcomes. The peak structure in $\zeta$, which is interpreting the maximum
breaking of conformal symmetry, is basically originated from two important quantities, 
associated with speed of sound and temperature dependent quark masses, which are adopted
in both approaches. However, two approaches contain two individual parameters, closely
related with spectral width of Kubo formalism, which inversely control the peak strength
of $\zeta$. We have tried to observe their role for equivalence in $\zeta$.
At low temperature domain ($T<T_c$), we have noticed a numerical discrepancy 
in $\zeta(T)$ for these two approaches. The main reason is that the RTA expression
contains the effect of folding by (PNJL) distribution function of quark, while it
is absent in LET expression. 

Our finite density results of bulk viscosity are indirectly reflecting the 
fact of QCD phase diagram - the transition occurs at lower temperature as
quark chemical potential becomes non-zero and increases. This can be realized
from the shifting of peak position in $\zeta$ towards the lower temperature,
when we increase quark chemical potential.

\section*{Acknowledgments}
KS and SU would like to acknowledge Council of Scientific and Industrial Research (CSIR)
 and Department of Science and Technology (DST) for funding this work. SU would like to 
thank Avik Banerjee for useful suggestions.
SG is financially supported from UGC Dr. D. S. Kothari Post Doctoral Fellowship under
grant No. F.4-2/2006 (BSR)/PH/15-16/0060.

%\section*{References}
%
%They are to be cited in the text in superscript
%after the punctuation marks e.g.~word,\cite{Marnelius} and word:\cite{Bjorken}
%If it is mentioned in the text as part of a sentence, it should be of normal size,
%e.g.~see Ref.~\refcite{Bohr}. Please list using the style shown in the following examples.
%For journal names, use the standard abbreviations. Typeset references in 9 pt Times Roman.
%Each reference number should consist of one reference only.
%

\end{document}